# What *really* happened on September 15th 2008?
## Getting The Most from Your Personal Information with Memacs

Karl Voit
Institute for Software Technology (IST)
Graz University of Technology
PIMresearch@Karl-Voit.at

## ABSTRACT
Combining and summarizing meta-data from various kinds of data sources is one possible solution to the data fragmentation we are suffering from. Multiple projects have addressed this issue already. This paper presents a new approach named *Memacs*. It automatically generates a detailed linked diary of our digital artifacts scattered across local files of multiple formats as well as data silos of the internet. Being elegantly simple and open, Memacs uses already existing visualization features of GNU Emacs and Org-mode to provide a promising platform for life-logging, Quantified Self movement, and people looking for advanced Personal Information Management (PIM) in general.

## 1. INTRODUCTION
In recent times, there is an increasing trend towards awareness of data in many kinds of ways. People who are engaged in lifelogging collect all kinds of data: digital artifacts they get in touch with like files, emails, and tasks. They also try to get as much data as possible from the physical world into their digital world: digital photographs, physiological data (such as a record of their heart beat rate), the amount of coffee they consume, the frequency of doing certain habits, and so forth.

Quantified Self is a very committed community that processes, combines, and visualizes this data collection. That results in deeper insight into their habits, behavior, and life. Direct feedback makes a feedback loop possible. New Year's resolutions like reducing the number of cigarettes smoked or drinking less cups of coffee are monitored objectively. As a result, people actually *see* their progress which leads to a more disciplined and encouraged attitude.

Therefore, collecting data is getting more and more important for a larger number of people. Unfortunately, this is not an easy task because our digital world is complicated. We do not only have data distributed among multiple devices like desktop PC, notebooks, smartphones, and mobile/external storage systems. Furthermore, we produce data that resides outside of our direct area of influence. Our digital artifacts are scattered across chat rooms, web forums, web-based blogs, authoring systems, server-based version control systems, and further web services of various kinds.

It is obvious that our digital artifacts are not solely stored in files. We have to use application programming interfaces (API) to access chunks of data, delivered via different kinds of formats such as XML, JSON, or even binary formats.

Therefore, we clearly need tools that help us to re-collect the data we produce outside of our computers, data that belongs to us, data we want to store locally. This is the only possibility to be able to control an all-embracing backup, process the data, and to get the most of our digital property.

## 2. RELATED WORK
In 1945, Vannevar Bush wrote "As We May Think" [2] where he presented a hypothetical thing called *Memex* which "is a device in which an individual stores all his books, records, and communications, and which is mechanized so that it may be consulted with exceeding speed and flexibility." Although Bush was thinking of a huge storage within a desk containing microfilms that is operated with levers, his vision was enormously visionary. He described many features and tools which were only introduced several decades later.

The aim of the "Lifestreams" project [9, 10] was to help users in optimizing their effort and time while managing local files and events. They tried to increase their ability to re-find and make use of information. Lifestreams used a unified time-oriented visualization metaphor.

Microsoft Research worked on a project which was later called "MyLifeBits" [11, 1]. Trying to implement a modern Memex, they integrated common Microsoft software tools for collecting local computer files, photographs, sounds, emails, appointments, web history, phone calls, and instant messaging into a database. Using Gordon Bell as a test subject, they digitized his physical artifacts such as personal paper documents (like notes, public and health records), books, and audio CDs. They also developed a neck worn digital camera called SenseCam, which was equipped with sensors that could trigger photos when something interesting happens. This allowed extensive capturing and archiving of visual impressions from a first person view.

The "Rememberance Agent" [13, 15, 14] represented another approach. Besides extracting information using the SMART information retrieval program, it presented the user relevant links to previously seen local files within the GNU Emacs editor. This way, the user got an assistant tool that displays helpful data associated to the user's current context.

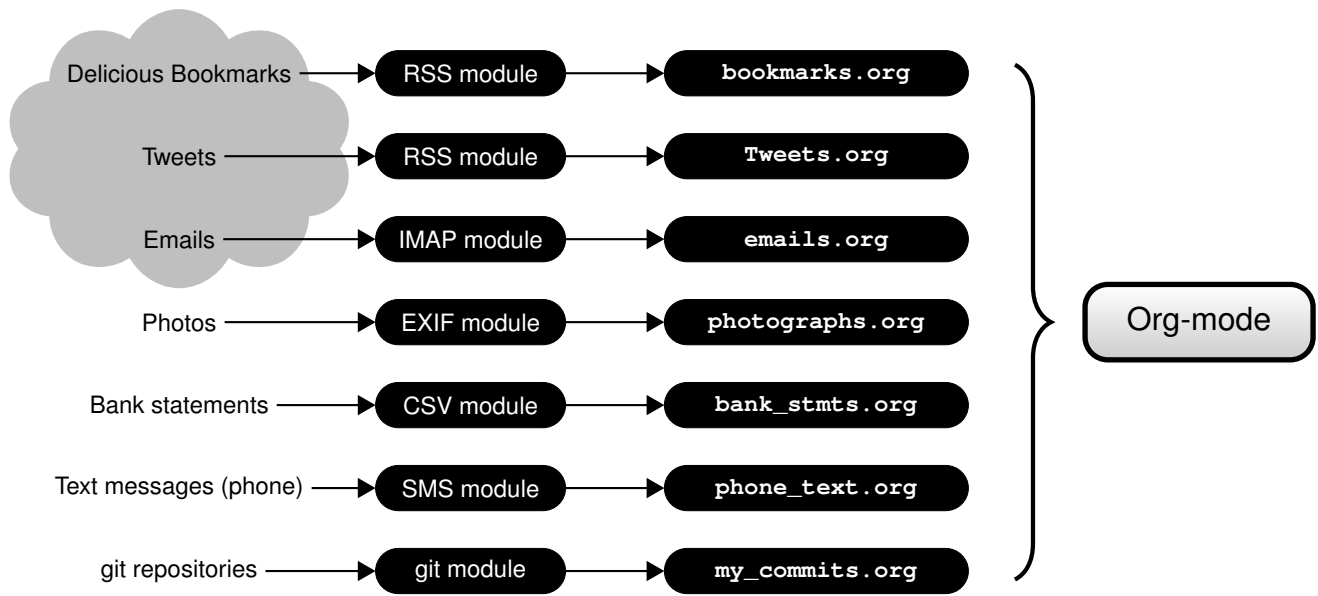

Figure 1: An example Memacs schema: On the left hand side there is a selection of web-based and local data sources. The black colored elements are part of *Memacs*: data-source specific modules are transcribing meta-data into Org-mode format files which are being interpreted, visualized, and handled by Org-mode.

"Presto" [6, 7, 5] combined automatically derived meta-data of collected documents with manually user-defined properties in a clever data structure. Within its graphical search interface called "Vista", the user could define dynamically updated sets using those properties. For this reason, Presto took advantage of many sources of meta-data the user can use to re-find local documents.

The "Haystack" project [12] provided a similar search interface. But instead of extracting meta-data of files, Haystack addressed a different layer. Arbitrary information objects could be linked using Resource Description Framework (RDF). This provided links between all kinds of data objects: starting from a document, the user was able to browse to the contact information of the author and further documents, emails, or links associated to that person.

"Stuff I've Seen" [8] (SIS) used a so called "Gatherer module" to extract, analyze, and combine data from local files, Outlook emails, Internet Explorer history and bookmarks. In combination with four other modules, SIS offered a decent local search engine combined with its own browser for local data. When presenting search results, the interface placed more recent items above older items. This resembles a kind of a temporal view.

## 3. MEMACS

*Memacs*[1] (a combination of "Memex" and "Emacs"[2]) is a system comparable to Memex, MyLifeBits, Lifestreams, and so forth. Unlike those projects mentioned, Memacs is based on open source components, is very easily set up, can integrate a bigger number of data sources, and provides additional benefit to the user with no additional cost or effort (besides the one-time setup process).

The Emacs platform was chosen because Org-mode is a very capable, extendable PIM framework. It did only require surprisingly small implementation effort in order to get functionality similar and even more advanced than huge funded research projects mentioned in the previous section.

Memacs consists of two parts: a set of Memacs-modules and visualizations done by Org-mode.

### 3.1 Org-mode
Org-mode[3] is an extension to the Emacs editor as Memacs is an extension to Org-mode. It is an open source project which enables users to accomplish a very large number of things: starting with basic outlining methods, it does provide headings, lists, features from Wiki systems, meta-data management systems, task and project planning, hypertext systems, tagging, workflow management, spreadsheet processing, and many more.

Based on text files, it represents a future-proof and platform independent system. Because of its flexibility to include external data and programs, it resembles a viable tool for reproducible research [3] and literate programming [16, 17]. Its data is freely parsable, searchable, and new content can easily be generated by only sticking to a simple syntax comparable to a Wiki system.

Users are able to add and maintain data in form of entities of text completed with internal and external links. This resembles a similar approach like the *Data Soup* of the Apple Newton MessagePad

---

[1] https://github.com/novoid/Memacs/
[2] (GNU) Emacs is one of the most powerful text editor systems: http://www.gnu.org/software/emacs/
[3] http://orgmode.org and documented in [4]

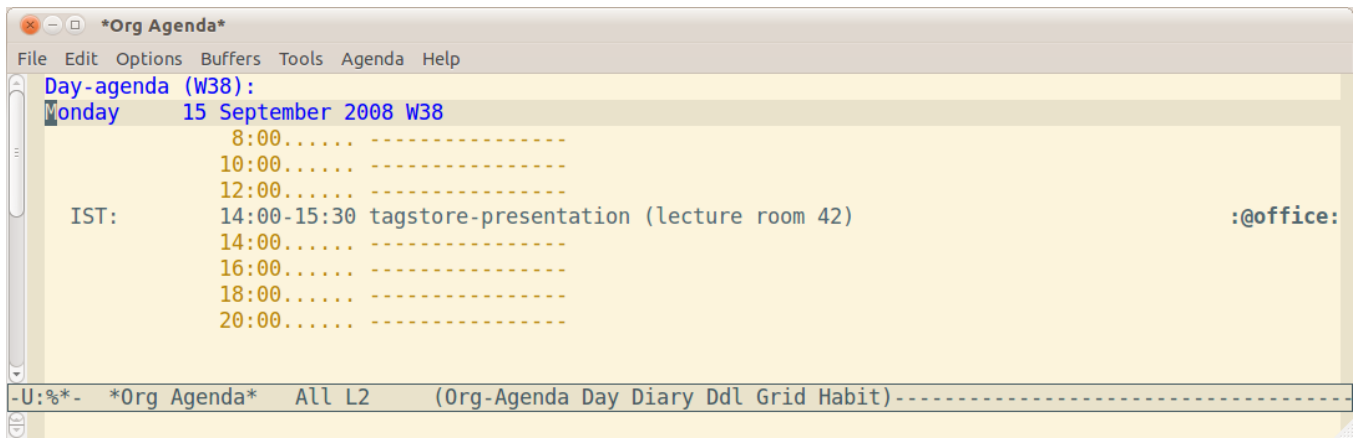

Figure 2: An example Org-mode agenda showing a single event on September 15[th] 2008 like any other usual personal information management (PIM) software would show.

PDA [18] or the Haystack project [12]. Org-mode files contain a mixture of notes, events, tasks, projects, archived data, references, tables, formulas, and links to other data entities or external files/resources.

### 3.2 Memacs Modules

*Memacs modules* generate, extract, and/or summarize data from external data sources and produce text-files in Org-mode syntax as shown in Figure 1. Resulting files are included into the Org-mode configuration so that, for example, Org-mode Agenda view visualizes all data related to a specific day- or time-stamp. Thus, the purpose of Memacs modules is similar to *Agents* (Haystack project) or *Gatherers* (Stuff I've Seen).

Different sources are being polled for time-related information of all kinds. This way, for example digital photographs appear on the Org-mode Agenda (calendar view) according to the second they were shot using a digital camera.

To avoid unnecessary duplication, only the meta-data is held in Memacs result files. Depending on the type of the data source, there will be a time-stamp of the information, a short summary, tags, and – most important – a link to the original data.

There are Memacs modules for a wide variety of data: timestamps in filenames[4], emails (IMAP, POP3, and maildir), RSS-feeds[5] for a large number of data feeds[6], versioning systems like git[7], calendars using iCalendar-format[8], generic Comma-Separated Values (CSV) files[9], generic (XML) files, exchangeable image file format (EXIF)[10], smartphone text messages, smartphone phone calls, and newsgroup postings.

Due to an open source license and a very modular design, new modules can be implemented with minimal effort. For example the personal information management (PIM) research software *tagstore*[11] [20, 21, 19] maintains a text log file. This log file can be easily parsed for file storage activities which automatically appear in the Org-mode agenda.

### 3.3 Visualization

Org-mode provides a wide variety of advanced data visualization features. *Folding* hides textual details and shows only the main structure of a document, the top level headings. By applying the unfolding command on a given heading, the user expands the next layers of headings and/or the complete sub-tree of information.

For showing matching headings and content related to a given search string or even tags, Org-mode generates so called *Sparse Trees*. Large documents can easily be scanned for relevant occurrences.

As shown in Figure 2 and Figure 3, Org-mode can derive temporal views for days, weeks, months, and other arbitrary number of days. It scans all Org-mode files and summarizes time-relevant data into one combined view called *Agenda*. The agenda can also be filtered by tags, priorities, text strings, and other meta-data.

Those internal visualization methods are easily adapted to specific requirements. For even more elaborated visualizations, Org-mode offers a broad range of export mechanisms such as plain text, HTML, Portable Document Format (PDF), LaTeX, DocBook, OpenDocument Format (ODF), Extensible Open XHTML Outlines[12] (XOXO), iCalendar, or even project management tools like TaskJuggler[13].

The active community of Org-mode is constantly developing more features for visualization, import, and export. With little effort, bidirectional synchronization solutions can be developed to integrate with other systems like the Google online services and so forth. This way, Google Calendar can be integrated into Org-mode

---

[4] file name conventions like `2012-04-13_Document.txt` or `2012-04-13T15.29_Image.jpeg` (based on ISO 8601)
[5] Resource Description Framework (RDF) Site Summary
[6] for example: Twitter `http://twitter.com`, social bookmark service `http://delicious.com`, news portals, event services, dynamic web notifiers, and much more
[7] git `http://git-scm.com/` or Apache Subversion `http://subversion.apache.org/`
[8] `http://tools.ietf.org/html/rfc5545`
[9] for bank statements and many more data which resides in this popular and generic file format
[10] used as meta-data storage format in most digital cameras
[11] `http://tagstore.org`
[12] `http://microformats.org/wiki/xoxo`
[13] `http://www.taskjuggler.org/`

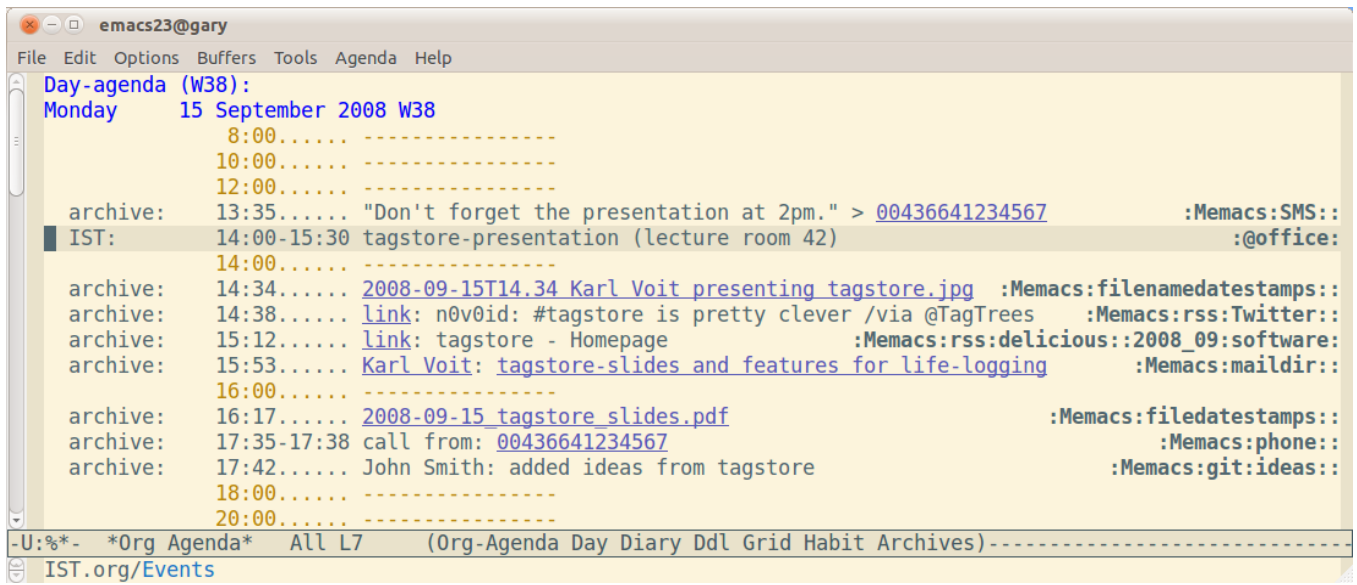

Figure 3: An example Org-mode agenda including meta-data from *Memacs*. This is the same day as in Figure 2 but with much more context information about digital artifacts that were created by the user. That way, the user re-accesses data that relates to the presentation to a much greater extent.

using open standard APIs.

### 3.4 Implementation and Setup

Memacs is implemented in Python 2.7 using various external libraries to adopt several data sources. The project is under an open license[14] and publicly hosted on github[15].

New Memacs modules are easily written since shared behavior is moved to common class files.

The user sets up a Memacs module only one time. This is a process which should not take longer than a couple of minutes. Periodically polling of data sources is done using usual mechanisms provided by the operating systems[16]. Once the system starts polling, the summarized meta-data appears automatically within Org-mode as long as the module data interface does not change.

There are two basic modes, a Memacs module can run in: overwrite mode and append mode. Services like RSS usually provide only the last twenty items. Such modules should be run in append mode where already known data is recognized (and ignored) and new data is appended. Other data sources like file-name datestamps can be run in overwrite mode since complete re-creation uses less processing power than comparing. Some Memacs modules offer both modes, other modules offer only one mode they can be used in.

In case of an error, the user is notified via Org-mode entries in the output files. Problems therefore appear on the agenda of the user where they are recognized easily.

---

[14]GNU General Public License (GPL) version 3
[15]https://github.com/novoid/Memacs/
[16]Microsoft Windows: Windows Task Scheduler; Mac OS X: launchd; GNU/Linux: crond

### 4. DISCUSSION

The benefits of a system like Memacs are easier to understand when users see a real-world scenario using real-world data.

Let's assume, a user remembers a specific presentation of a research software called *tagstore*. Usual calendar software products hold a history of all events and offer a search function. Our example user searches for the string *tagstore* and is confronted with the day view of September 15[th] 2008, where she joined a presentation held by the project manager of *tagstore*. She gets a view more or less similar to Figure 2: one single event took place on this day. If the user did not note anything directly within this calendar entry, this is everything she gets as information about this past event.

Using a system like Memacs offers way more contextual information about the event. Figure 3 shows the same day as before but enriched with helpful meta-data. It starts with 1:35 pm where she sent a text message to a colleague, reminding him to join the presentation as well. Following the link (the phone number) leads to her address book (also maintained within Org-mode) and to the entry of her colleague John Smith. Therefore she can assume, that John Smith was also attending the presentation, probably having additional important knowledge or data related to that topic.

During the presentation at 2:34 pm, the filename-datestamp module found an image file representing a photograph. According to the file name, this snapshot shows the presenting person. This is a crucial mental support for people with difficulties in remembering faces or names. It is worth noting that the original file name (chosen by the camera device) does not even have to be changed manually: the temporal context within Memacs allows the user to conclude, what content the photograph[17] might offer. Most likely there will be photographs of single slides, the presenting person, the audience, or something else related to that event.

---

[17]with an unmodified generic file name from a camera like `IMG0042.jpg` (EXIF module) or with an time-stamp added like `2008-09-15T14.34_IMG0042.jpg` (filenametimestamp module)

At 2:38 pm, the user posted a comment on the micro-blogging platform Twitter[18]. Though she was tweeting the note for her contacts on Twitter back in 2008, she now benefits from her own message a couple years later. Otherwise she probably would not remember what she was thinking of the event. Half an hour later, she created a public visible bookmark on the social bookmark platform delicious[19]. This link most likely is handy for her re-investigation on the topic as well. Corresponding tags of the delicious entry ("2008_09" and "software") are transcribed by Memacs into native Org-mode tags.

Also relevant is the email, she got at 3:53 pm from the original author of the presentation: the subject line suggests that it included the slides shown and additional descriptions. By clicking on the subject line, she can retrieve the original email content. At 4:17 pm she saved the presentation slides to her file system. Once again: a click on the blue underlined section provides direct access to the PDF file.

The three minute phone call from 5:35 pm is an additional clue, what action she was taking related to the presentation. Especially since another three minutes later, John Smith (the person she was talking to on the phone) committed to the git repository using a summary that reflects ideas about the *tagstore* project.

This set of context-related meta-data overview gives the user the possibility to get the most of her information management without any additional manual effort. She can retrieve files, opinions, connections to other people, communication patterns and much more.

As a positive side effect, she got a copy of many important "cloud-based" information on her own, allowing independent backup and access.

## 5. CONCLUDING REMARKS

Memacs collects chunks of meta-data from a variety of local and remote data sources using Memacs modules. This automatism runs in background, with no need for additional manual effort for classification, description, and so on.

Once a user has set up a system like Memacs using minimal effort, huge benefits are noticeable in every day work. Using multiple visualization features from Org-mode, Memacs data gives a more complete impression, on the context of digital artifacts. The temporal view of the *Agenda* is just one of many ways of being able to access and browse the data managed within Org-mode. Filter mechanisms and other derived views of Org-mode or even external processing in other tools provide easy and fast ways to integrate, handle, visualize, or export data.

---

[18] http://twitter.com
[19] http://delicious.com